%% file: lic.tex
\documentclass[lic,lith,english]{liuthesis}
\include{settings}

\department{Institutionen för teknik och naturvetenskap}
\departmentenglish{Department of Science and Technology}
\edition{1:1}

\titleenglish{A Digitalization Framework for Smart Maintenance of \\Historic Buildings}
\division{Division of Physics, Electronics and Mathematics}
\thesisnumber{1973}

\publicationyear{2023}
\isbnprint{8075}{304}{3}
\isbnpdf{8075}{305}{0}
\author{Zhongjun Ni}

\begin{document}

\chapterstyle{VZ43}


\include{intro}

\printbibliography[title={References}]


\end{document}

%% file: settings.tex
\usepackage[backend=biber,style=ieee,doi=false,hyperref]{biblatex}
\ifxetex 
\fi

\usepackage{amsmath,amssymb,amsfonts}
\usepackage{algorithmic}
\usepackage{graphicx}
\usepackage{booktabs}
\usepackage{multirow}
\usepackage{listings}
\usepackage{calc,color}
\newif\ifNoChapNumber

\makeatletter
\makechapterstyle{VZ43}{

}
\makeatother



%% file: intro.tex

\chapter{Introduction} 
\label{cha:introduction}

Digital transformation has become a prominent phenomenon in the modern era, revolutionizing all aspects of society and industry. As valuable cultural assets, historic buildings need careful maintenance to keep their functionalities and heritage values. This chapter introduces the implications, benefits, and challenges of digital transformation in maintenance of historic buildings.

\section{Background} 
\label{sec:background}

This section introduces digital transformation and its potential impact on maintenance of historic buildings. It starts with an overview of digital transformation, where key concepts and enabling technologies are introduced. Then, the practical needs and requirements for maintenance of historic buildings are discussed. Among maintenance strategies, preventive approaches allow for managing risks while controlling costs. Finally, the role of digital technologies in real-time monitoring, data analytics, and documentation is highlighted.

\subsection{Overview of Digital Transformation}

Digital transformation is a process that attempts to improve an entity by using various digital technologies to cause considerable changes to its attributes~\cite{vial_understanding_2019}. For example, digital transformation in the industry often involves dramatically rethinking an organization's use of technology, culture, human resources, and processes to improve business performance~\cite{nath_industrial_2020}. Proper digital transformation can bring substantial economic benefits. For instance, it might unlock an additional US\$1.25 trillion market valuation for all Fortune 500 companies~\cite{smith_unleashing_2023}. Driven by those advantages, spending on digital transformation has grown steadily in recent years. As the International Data Corporation reported in 2022~\cite{simpson_worldwide_2022}, global spending on digital transformation is estimated to grow at a compound annual growth rate of 16.3\% over the next five years and reach US\$3.4 trillion by 2026.

In digital transformation, various digital technologies, such as information, communication, networking, and computing, play a central role. Among them, two representative technologies with significant impact are Internet of Things (IoT) and cloud computing. IoT enables stakeholders to monitor interested entities or processes in real-time through widely deployed sensors, actuators, and embedded communication hardware~\cite{gubbi_internet_2013}. Whereas IoT facilitates data collection, cloud computing provides powerful storage and computing capabilities to process collected massive amount of data~\cite{liu_methodology_2021}. The emergence of public cloud computing platforms, e.g., Amazon Web Services~\cite{amazon_aws_2021}, Microsoft Azure Cloud~\cite{microsoft_azure_2021}, and Google Cloud~\cite{google_cloud_2021}, enables organizations and customers to access applications on demand from any place by providing end-to-end service provisioning.

As more and more data are available, it becomes critical to analyze them to provide applications or services to users. Machine learning and data mining methods are useful to discover knowledge from data and further develop advanced artificial intelligence (AI) applications, such as forecasting~\cite{zhang_social_2021} and anomaly detection~\cite{liu_anomaly_2020}. Recently, deep learning methods, a subset of machine learning that is represented by various artificial neural networks, have become popular in data analytics due to their improved capabilities in dealing with massive volumes of data, extracting features, and modeling nonlinear processes~\cite{runge_review_2021}. Open-source machine learning frameworks, e.g., TensorFlow~\cite{abadi_tensorflow_2016} and PyTorch~\cite{paszke_pytorch_2019}, have also greatly simplified network implementation and model training.

Thanks to real-time monitoring and advanced data analytics, the concept of digital twins is increasingly popular in digital transformation. A digital twin is a complete physical and functional modeling of a component, product, or system that includes all necessary information that might be used in the present and following lifespan stages~\cite{boschert_digital_2016}. Creating a digital twin is useful when an interested physical entity changes over time. Digital twins could provide complete support for decision-making across diverse operations by assessing current situations, diagnosing previous issues, and forecasting future trends~\cite{qi_enabling_2021}.
The application of digital twins has been witnessed in sectors like utilities~\cite{wang_survey_2022}, oil and gas production~\cite{bp_twin_2018}, manufacturing~\cite{glaessgen_digital_2012, tao_digital_shop_2017}, and health services~\cite{liu_novel_2019}.

\subsection{Maintenance of Historic Buildings}

Historic buildings are generally considered to be buildings or structures that have a historic value. The historic value might be reflected in the building itself, such as its design, construction methods, and architectural significance. It could also be associated with a person of national significance or be significant for a specific historical event or period~\cite{historic_2022}. Historic buildings must be protected to preserve their heritage values so that future generations can understand and appreciate humanity's diverse cultures and history~\cite{ashrae_handbook_hvac_applications_2011}. However, historic buildings are vulnerable to deterioration induced by a variety of factors, including physical, chemical, natural, as well as human actions. For instance, wooden items and structural elements are susceptible to dramatic fluctuations in ambient relative humidity (RH)~\cite{en_15757_conservation_2010}. Those risks should be mitigated with good maintenance.

Maintenance of a building involves all operations performed during the service life of a building or its components to keep them in a state where they can perform their required functions~\cite{iso_15686_1_buildings_2011}. Maintenance can be generally classified into two types: planned and unplanned. Planned maintenance is organized, controlled, and follows a known procedure. It aims to maximize the performance of components and prevent problems that can emerge predictably during their lifetime~\cite{mishra_maintenance_2012}. In contrast, unplanned maintenance occurs without a plan. Each action is performed in response to a problem, potentially resulting in prolonged breakdowns, poor user experience, and loss of control~\cite{mishra_maintenance_2012}.

Planned maintenance is a smarter way to preserve historic buildings because it enables managing risks before they cause damage. To achieve planned maintenance, it is effective to take preventive approaches. With preventive maintenance, more costly and risky extensive restoration operations can be avoided by identifying and repairing minor deterioration in historic buildings~\cite{eken_evaluation_2019}. This approach to maintenance is also consistent with the reality of building conservation, which is to manage risk rather than avoid it entirely~\cite{ashrae_handbook_hvac_applications_2011}. For instance, many historic buildings are still used for public purposes today, e.g., as museums, libraries, theatres, and religious sites. For such buildings, one priority of preventive maintenance is to create an appropriate indoor environment to ensure preservation conditions~\cite{lucchi_review_2018} while considering the human comfort required by public access and sustainability requirements, such as energy efficiency.

\subsection{Role of Digital Technologies in Maintenance of Historic Buildings}

Digital transformation is reshaping the built environment by driving efficiency, sustainability, and innovation throughout the life cycle of buildings and infrastructure. Maintenance of historic buildings is an exciting scenario for the application of intelligent digital solutions because each historic building is unique due to factors such as built era, architectural style, construction method, and purpose of use. Smart maintenance of historic buildings benefits from real-time monitoring, advanced data analytics, and documentation.

Real-time monitoring enables facility managers to know the most recent state of a building and make timely interventions when necessary~\cite{colace_iot_2021}. It involves collecting information on indoor environment and building infrastructure in several subsystems like electric power, lighting, water, as well as heating, ventilation, and air conditioning (HVAC) systems~\cite{balaji_brick_2018}. Real-time monitoring can be achieved by deploying IoT devices to collect data and utilizing visualization tools to display the collected data user-friendly.

The potential of collected data can be further explored through analytics to provide deeper insights. Data analytics benefit the maintenance of historic buildings in several aspects, including indoor environment control, facility management, and energy optimization~\cite{jia_adopting_2019}. For example, a public historic building can be operated according to specific needs, such as adaptively scheduling the HVAC and lighting systems based on occupancy levels. In facility management, data mining methods allow for evaluating the operating conditions of building components and predicting the occurrence of degradation, malfunction, or breakdown~\cite{arditi_issues_1999}. This predictive approach enables timely interventions while avoiding interference with ongoing core activities. Energy efficiency in historic buildings is a critical issue for meeting global sustainability. Analyzing the energy consumption data of a building allows for identifying patterns and seeking the potential for energy savings.

Documenting historic buildings includes three dimensional (3D) modeling of geometry as well as semantic knowledge information management. Many technologies, such as photogrammetry, 3D computer graphics, laser scanning, and ontology, are currently used for architectural heritage modeling and management~\cite{yang_review_2020}. With 3D geometry of a historic building as well as its components' attached attribute, material, and connection information, an information model can thus be created. As-built building information modeling (BIM) is increasingly being used to model, manage, and restore historic buildings~\cite{cogima_role_2019}. Since the information stored in BIM models is usually static, integrating IoT and BIM to update the information automatically is efficient in reducing manual labor~\cite{rosati_air_2020}.

\section{Motivation} 
\label{sec:motivation}

Previous studies have used digital technologies for monitoring, remotely operating, and modeling historic buildings. However, some research gaps and challenges still exist, including the resources needed to handle the ever-expanding collected data, organizing these heterogeneous data with a consistent data format, and leveraging the latest data analysis methods to analyze the data to optimize maintenance of historic buildings.

Digitalization of historic buildings calls for a solution with high availability and scalability, because conservation of historic buildings is a long-term process. In this process, the volume of collected data continues to grow, thus requiring storage and computing resources to store and analyze those data to extract insights for guiding maintenance. Previous studies~\cite{huynh_wireless_2010,zhang_remote_2010,guo_ima_2012,lazarescu_design_2013,corbellini_cloud_2018} mainly used private servers to handle collected data. However, private servers need ongoing maintenance and support, which brings considerable cost and accountability. Additionally, scaling up private servers to handle the increase in demand requires a large upfront investment. Therefore, there is a critical need for a digitalization solution that can provide highly available and scalable storage and computing resources to meet the long-term preservation needs of historic buildings.

Organizing collected data from historic buildings to create parametric digital twins is a challenging task due to the diversity of data sources. Data could come from analytical studies, diagnostics, and monitoring~\cite{bruno_historic_2018}, which are generated by various models, methods, and tools utilized by specialists with various backgrounds~\cite{acierno_architectural_2017}, such as facility managers, building engineers, and conservation researchers. Furthermore, historic buildings might be equipped with various building management systems (BMS). Even the most advanced BMS generates data flows that vary between buildings, locations, and vendors~\cite{balaji_brick_2018}. As a result, it is critical to use a uniform, reusable, and expandable data format to represent data about historic buildings. Lacking such a representation makes it difficult to extend a solution developed for one historic building to other buildings. 

The latest data analysis methods, e.g., deep learning, have not been fully employed to optimize maintenance of historic buildings. For example, improving energy efficiency is one aspect of maintenance optimization. In the EU, $\sim$35\% of buildings were built 50 years ago, and $\sim$75\% of them are energy inefficient~\cite{buildup_energy_2019}. Deep learning methods can be leveraged to understand energy consumption patterns of historic buildings and develop applications such as energy consumption forecasting, HVAC optimization, as well as fault diagnosis and detection~\cite{somu_deep_2021}.

\section{Objectives} 
\label{sec:objectives}

This thesis set out to develop a digitalization framework to improve maintenance of historic buildings. The framework is based on public cloud services, thus providing secure, robust, and scalable storage and computing resources for handling data collected by IoT devices as well as from other systems in buildings. Ontology is used to represent these data from diverse sources in a consistent data format to create parametric digital twins and facilitate subsequent data analysis tasks. Because much of the data collected in buildings are time series, the latest data mining methods, such as deep learning, are used to analyze these data to provide insights.

Based on the framework, it is expected to offer a better knowledge of the operational and usage circumstances of historic buildings and make it possible to achieve smart maintenance, such as preventive conservation and predictive control. To achieve this goal, the following sub-goals G1--G3 are addressed in this thesis:
\begin{itemize}
\item[\textbf{G1:}] Develop a cloud-based IoT system by integrating easy-to-use hardware, open-source software libraries, and public cloud services for end-to-end data collection, transmission, storage, processing, and visualization.
\item[\textbf{G2:}] Formulate a solution for creating parametric digital twins of historic buildings that can reliably represent data about buildings, reflect the most recent operational state, and allow for subsequent data analytics.
\item[\textbf{G3:}] Construct advanced data analysis approaches, e.g., deep learning-based, to discover insights from data collected from historic buildings, like indoor environment and energy use, to optimize maintenance and achieve conservation.
\end{itemize}

These sub-goals are achieved in the included papers as illustrated in Table~\ref{tab:subgoals}. \textbf{Papers I} and \textbf{II} focus on concept formulation and preliminary validation. In \textbf{Paper I}, the methodology was clarified for integrating multiple digital technologies to create digital twins. \textbf{Paper II} presents a reference implementation of the framework. The implemented solution is a complete IoT system for data collection using various environmental sensors, data transmission through an edge platform, and data storage with the Microsoft Azure Cloud.

\begin{table}[!htb]
\small
\centering
\caption{Focus of each paper in achieving the sub-goals.}
\label{tab:subgoals}
\begin{tabular}{@{}lccc@{}}
\toprule
Papers & G1 & G2 & G3 \\ \midrule
I & \multicolumn{2}{c}{\multirow{2}{*}{\begin{tabular}[c]{@{}c@{}}Concept formulation\\ Preliminary validation\end{tabular}}} &  \\
II & \multicolumn{2}{c}{} &  \\
III & Field test &  &  \\
IV &  & Field test & Analysis of indoor environment \\
V &  &  & Multi-horizon building energy forecasting \\
VI &  &  & One-step-ahead building energy forecasting \\ \bottomrule
\end{tabular}
\end{table}

\textbf{Paper III} describes a field test that was conducted to verify the developed system in three historic buildings, namely the City Theatre, the City Museum, and the Auditorium, in Norrk\"oping, Sweden. The field test verifies the stability of the system regarding the long-term operation for data collection. In \textbf{Paper IV}, ontology was further introduced into the solution to provide a consistent schema for organizing and representing data of historic buildings. A case study conducted in the City Theatre validates the feasibility of the solution, assesses the influence of occupants' presence on indoor environment, as well as identifies potential risks. 

In \textbf{Papers V} and \textbf{VI}, the performance of cutting-edge deep learning methods was investigated in making both point and probabilistic forecasts of building energy consumption. \textbf{Paper V} focuses on multi-horizon prediction. Model performance was examined in the City Theatre and the City Museum, which have different operating modes. \textbf{Paper VI} concentrates on one-step-ahead prediction and illustrates the combination of predictive models with a digital twin model to improve building energy performance.

\section{Thesis Outline} 

The remainder of this thesis is organized as follows. Chapter~\ref{cha:methodology} describes the research methodology, including essential modules of the digitalization framework, indoor environment monitoring, and data-driven building energy forecasting. Then, Chapter~\ref{cha:digitalization_framework} introduces the detailed design of the framework and field test in three case study buildings. After that, a summary of included papers together with author's contributions are given in Chapter~\ref{cha:summary_of_included_papers}. At last, Chapter~\ref{cha:conclusion_and_future_work} concludes the thesis and identifies future work in the remaining doctoral studies. Chapters 6--11 are included papers.

\chapter{Methodology}
\label{cha:methodology}

This chapter first describes essential modules of the digitalization framework. It then introduces two representative applications for smart maintenance: indoor environment monitoring and building energy forecasting. Overall, this research focuses on applied research.

\section{Digitalization of Historic Buildings}

\begin{figure}[!ht]
\includegraphics{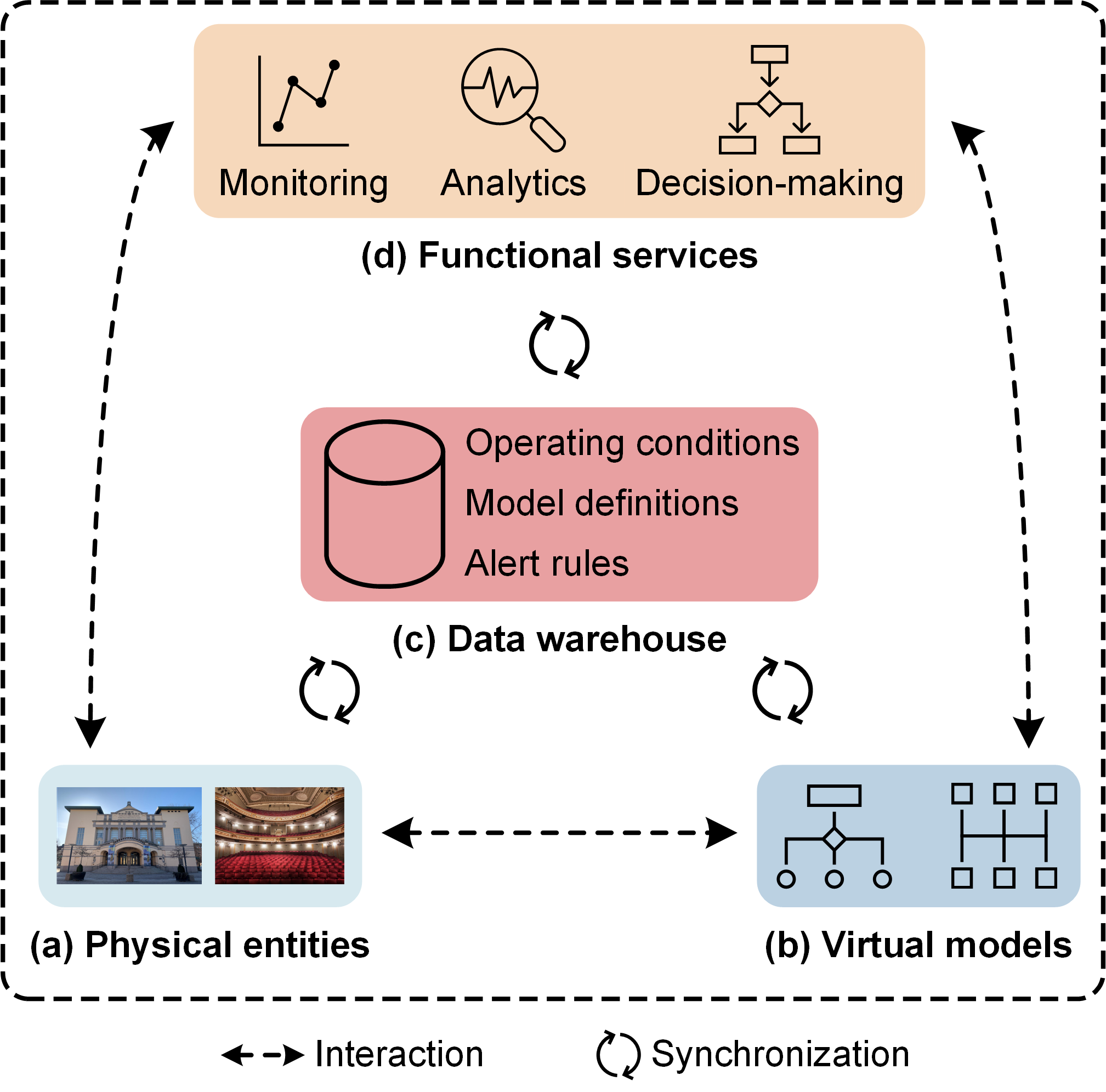}
\centering
\caption{Essential modules in the proposed digitalization framework~\cite{ni_enabling_2022}.}\label{fig:design_overview}
\end{figure}

Digitalization of historic buildings aims to provide a common operating picture by collecting, organizing, processing, and presenting data from multiple sources. Based on recommended designs of IoT systems~\cite{jia_adopting_2019, iso_30141_iot_2018} and the general needs of the maintenance of historic buildings, the digitalization framework consists of four essential modules: physical entities, virtual models, a data warehouse, and functional services. These modules communicate with one another in various ways, including command interaction and data synchronization (see Fig.~\ref{fig:design_overview}).

\subsection{Physical Entities}

Physical entities are real-world objects, including spaces and assets within those spaces. Spaces can range from an entire building to specific parts of a building, including sub-buildings, floors, or individual rooms. Assets, on the other hand, are things that are placed within a building but are not regarded to be part of its structure. Examples of assets include housed artifacts and infrastructures like water supply, lighting, as well as HVAC systems.

The physical entities to be involved depend on the purpose of an application. For instance, when focusing on the preservation of collections within an exhibition space, physical entities of interest may include the indoor environment of the space. On the other hand, if the aim is to optimize building energy use, physical entities to be considered could include energy-consuming equipment and spaces they serve.

\subsection{Virtual Models}

A virtual model is a digital representation of a physical entity, designed to capture and define its essential properties and behavior. Virtual models are able to alter in response to changes in the status of the represented physical entities. In addition, they should offer interfaces for other modules to notify and get model changes. These specifications ensure that virtual models have the required information to support subsequent tasks.

\subsection{Data Warehouse}

A data warehouse includes definitions of data structures, data storage, and operating procedures. It stores various data associated with physical entities, including operating conditions, model definitions, and alert rules. Operating conditions are the current and historical conditions of physical entities. Model definitions contain definitions of virtual models as well as data analysis models aimed at extracting insights from the data. Alert rules outline the actions for preventive conservation when specific conditions are met.

\subsection{Functional Services}

The goal of functional services is to empower users with valuable information and guiding them towards efficient and effective actions. These goals are met by integrating and analyzing data from multiple sources. The functional services can be broadly classified into three categories:
\begin{itemize}
    \item \textbf{Monitoring} includes the visualization of real-time and historical data of building operating status. Monitoring services enable effective tracking and oversight by providing users with a clear view of events and behaviors.
    \item \textbf{Analytics} employs techniques such as machine learning and exploratory data analysis (EDA) to derive insights from data. These insights help uncover patterns, trends, and correlations, enhancing the understanding of operating conditions.
    \item \textbf{Decision-making} uses the insights derived from analytics to guide building maintenance. These efforts include assessing current situations, diagnosing previous issues, and making informed decisions.
\end{itemize}

\subsection{Interactions and Synchronizations}

Interactions and synchronizations allow commands and data to be shared between modules. Interactions primarily concern command exchange, whereas synchronizations focus on data sharing. Typical interactions include:
\begin{itemize}
    \item Physical entities--Virtual models: When virtual models are manipulated, corresponding commands are sent to control the physical counterparts, facilitating timely adjustments or optimizations.
    \item Virtual models--Functional services: Functional services query virtual models to learn about the digital twin through its attributes and relationships.
    \item Data warehouse--Functional services: Functional services find relevant resources inside a building from the data warehouse. After that, the service performs some operations on the obtained resources.
\end{itemize}

Meanwhile, examples of synchronizations include:
\begin{itemize}
    \item Physical entities--Data warehouse: The data warehouse stores the most up-to-date status of physical entities on time.
    \item Virtual models--Data warehouse: Virtual models retrieve information from the data warehouse, enabling them to adapt and accurately reflect the status of their physical counterparts.
\end{itemize}

These interactions and synchronizations establish seamless connections between modules, enabling efficient data and command sharing as well as enhancing the overall functionality and effectiveness of the system.

\section{Indoor Environment Monitoring}

Indoor environment monitoring acts as a diagnostic tool for preserving historic buildings. It allows facility managers to know the operating status of a building in real time. In addition, analyzing collected environmental parameters helps to find possible risks that may harm building conservation or human comfort. Indoor environment monitoring serves two functions in this study. The first is to analyze the impact of occupants' presence on the indoor environment and the correlations between changes in various environmental parameters. The second is to evaluate RH fluctuations throughout the year to identify possible risks.

The influence of occupants on the indoor environment is investigated using a combination of EDA and classical statistical analysis. EDA is used to observe and compare changes in indoor environment and to investigate the factors that drive these changes. The influence of different occupancy levels on indoor environment is quantified using classical statistical analysis, such as one-way analysis of variance (ANOVA).

RH fluctuations are assessed using the method suggested in the European standard EN 15757:2010~\cite{en_15757_conservation_2010}. This standard is a reference for maintaining temperature and RH levels in historic buildings in order to avoid physical harm to organic and hygroscopic materials caused by climate change. 

\section{Building Energy Forecasting}

This study employs data-driven approaches to predict energy consumption of historic buildings. A data-driven energy forecasting model is built based on available data, such as past energy use, past outdoor weather, and future weather forecasts, to predict future energy use of a building. Both point and probabilistic forecasting are studied.

\subsection{Problem Formulation}

\begin{figure}[!ht]
\includegraphics{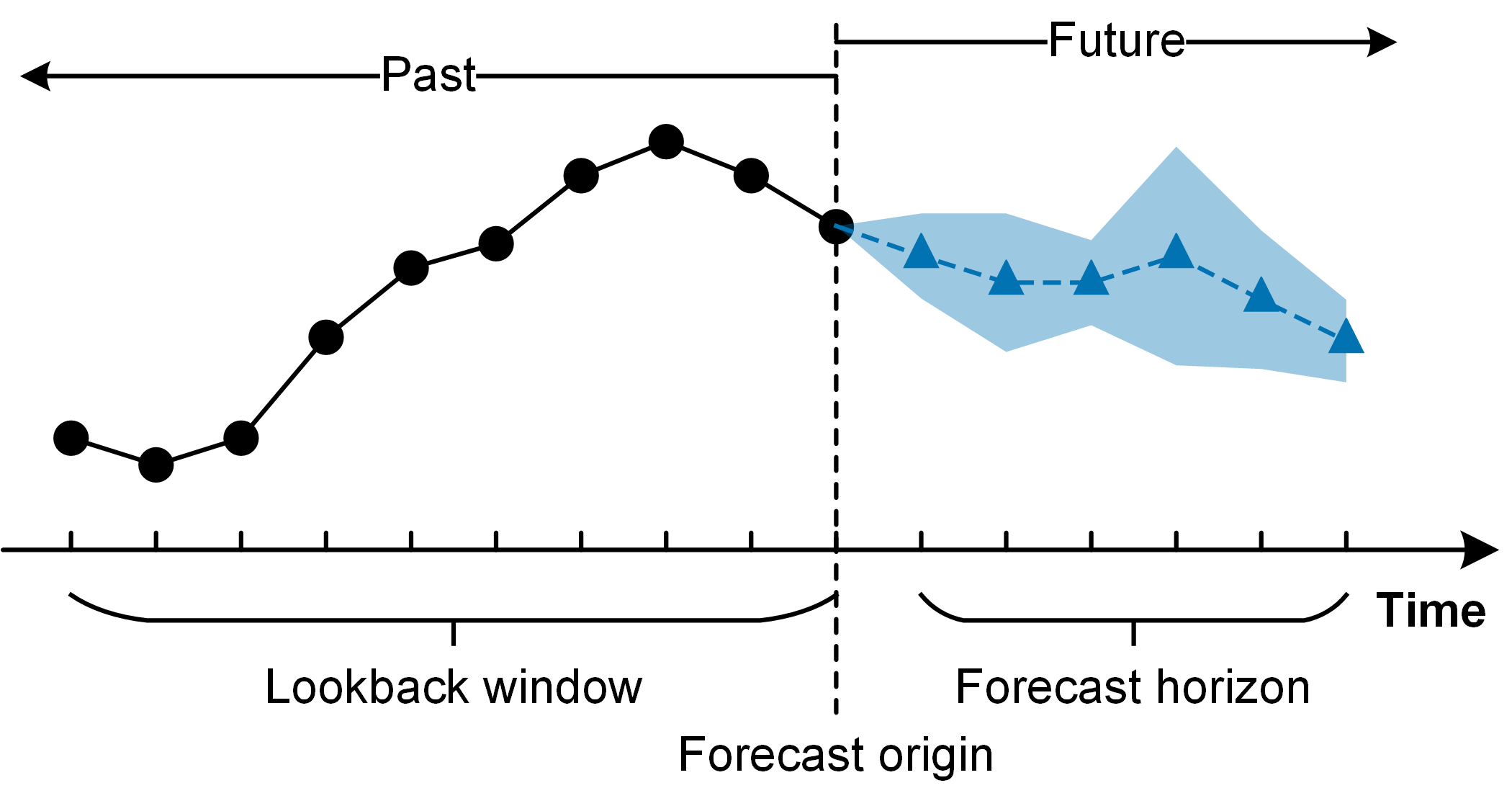}
\centering
\caption{An illustration of building energy forecasting. Black dots represent observed values of a type of energy use over a lookback window ($size=10$) in the past. Blue triangles refer to predicted values (conditional median or mean) of the energy use over a forecast horizon ($size=6$) in the future. The shadow area is a specific prediction interval~\cite{ni_study_2023}.}\label{fig:building_energy_forecasting}
\end{figure}

We denote a specific type of energy use as a non-negative real variable $y\in \mathbb{R}_{+}$. Factors that might affect the energy use are divided into two parts: observable in the past (i.e., before (including) a forecast origin, see Fig.~\ref{fig:building_energy_forecasting}), which is denoted as a real row vector $\mathbf{x_{b}} \in \mathbb{R}^{k}$ and observable in the future (i.e., after the forecast origin), which is denoted as a real row vector $\mathbf{x_{f}} \in \mathbb{R}^{m}$. All variables are assumed to be observed at fixed intervals and organized in a chronological order. The observed energy use at time $t$ is denoted as ${y}_{t}$. Likewise, observed affecting factors are denoted as $\mathbf{x_{b}}_{t} =[{x_{b}}_{1,t} ,{x_{b}}_{2,t} ,...,{x_{b}}_{k,t}]$ and $\mathbf{x_{f}}_{t} =[{x_{f}}_{1,t} ,{x_{f}}_{2,t} ,...,{x_{f}}_{m,t}]$, respectively.

Then, a point energy forecasting model takes the form
\begin{equation}
\hat{y}_{t+1:t+h} =f_{\theta }(y_{t-w+1:t} ,\mathbf{x_{b}{}}_{t-w+1:t} ,\mathbf{x_{f}{}}_{t+1:t+h}), \label{eqn:point_forecast}
\end{equation}
where $\hat{y}_{t+1:t+h} = \left[\hat{y}_{t+1},\hat{y}_{t+2},...,\hat{y}_{t+h}\right] \in \mathbb{R}_{+}^{h}$ are model forecasts for mean values of the energy use over a forecast horizon $h$, $y_{t-w+1:t} = [y_{t-w+1} ,y_{t-w+2} ,...,y_{t}] \in \mathbb{R}_{+}^{w}$ as well as $\mathbf{x_{b}}_{t-w+1:t} = \{\mathbf{x_{b}}_{t-w+1} ,\mathbf{x_{b}}_{t-w+2} ,...,\mathbf{x_{b}}_{t}\}$ are observations of the energy use and affecting factors over a loopback window $w$, $\mathbf{x_{f}}_{t+1:t+h} =\{\mathbf{x_{f}}_{t+1} ,\mathbf{x_{f}}_{t+2} ,...,\mathbf{x_{f}}_{t+h}\}$ are observations of affecting factors over the forecast horizon $h$, and $f_{\theta}{(.)}$ is the prediction function that the model has learned.

For developing probabilistic forecasting models, we do not assume that energy use of a building follows some distributions but develop models that directly generate interested quantiles through quantile regression~\cite{koenker_regression_1978}. The $p$th quantile denotes the value where the cumulative distribution function crosses $p$~\cite{hao_quantile_2007}. Therefore, quantiles can specify any position of a distribution.

Given a predetermined set of quantiles $\mathcal{Q} \subset (0,1)$, a quantile energy forecasting model takes the form
\begin{equation}
\hat{y}_{t+1:t+h}^{(p)} =g_{\theta}(y_{t-w+1:t} ,\mathbf{x_{b}{}}_{t-w+1:t} ,\mathbf{x_{f}{}}_{t+1:t+h}), \label{eqn:quantile_forecast}
\end{equation}
where $p$ is an element of the set $\mathcal{Q}$, $\hat{y}_{t+1:t+h}^{(p)} = \left[\hat{y}_{t+1}^{(p)} ,\hat{y}_{t+2}^{(p)} ,...,\hat{y}_{t+h}^{(p)}\right] \in \mathbb{R}_{+}^{h}$ are the model forecasts for the $p$th quantile of the energy use over a predicting horizon $h$, $y_{t-w+1:t}$, $\mathbf{x_{b}}_{t-w+1:t}$ and $\mathbf{x_{f}}_{t+1:t+h}$ have same definitions as in the point forecasting model, and $g_{\theta}{(.)}$ is the prediction function that the model has learned.

\subsection{Model Development}

\begin{figure}[!ht]
\includegraphics[width=\textwidth]{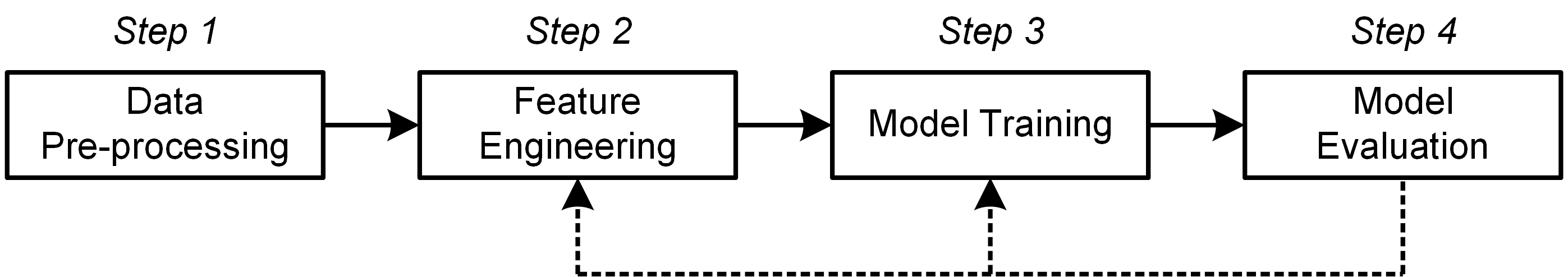}
\centering
\caption{Developing machine learning-based models involves four main steps~\cite{ni_improving_2021}. Data pre-processing usually includes detecting and correcting erroneous data, filling in incomplete data, and splitting data into three sets, namely training, validation, and test sets. Feature engineering is the process of extracting valuable features from raw data in order to improve model performance. Model training is selecting and employing effective machine learning algorithms for training models. When the training is over, the model is evaluated to see if its performance meets expectations.}\label{fig:machine_learning_model}
\end{figure}

Machine learning, especially deep learning, is used to develop building energy forecasting models. As shown in Fig.~\ref{fig:machine_learning_model}, the four main steps in developing a model are data pre-processing, feature engineering, model training, and model evaluation.

Point forecasting models were trained on a training set to minimize total squared error, which leads to forecasts of the mean~\cite{hyndman_forecasting_2018}. The training squared error for a set $\mathcal{S} =\{(y_{t-w+1:t} ,\mathbf{x_{b}{}}_{t-w+1:t} ,\mathbf{x_{f}{}}_{t+1:t+h} ,y_{t+1:t+h})\}_{t=w}^{n+w-1}$ is denoted by $L_{s}(\mathbf{\theta})$, and
\begin{equation}
L_{s}(\theta) = \sum _{t=w}^{n+w-1}\sum _{i=1}^{h} \left(\hat{y}_{t+i} -y_{t+i}\right)^{2}, \label{eqn:square_loss}
\end{equation}
where $n$ denotes the number of training samples and definitions of other variables are as in Eq.~\ref{eqn:point_forecast}.

Probabilistic forecasting models, i.e., quantile forecasting models in this study, were trained to minimize total quantile loss. As in studies~\cite{wen_multi-horizon_2017, lim_temporal_2021}, the $p$th quantile loss for one prediction at one time step is calculated as 
\begin{equation}
\ell(\hat{y} ,y,p)=(1-p)(\hat{y} -y)_{+} + p(y-\hat{y})_{+}, \label{eqn:quantile_loss}
\end{equation}
where $(.)_{+}=max(0,.)$. Then, the training quantile loss for a set $\mathcal{S} =\{(y_{t-w+1:t} ,\mathbf{x_{b}{}}_{t-w+1:t} ,\mathbf{x_{f}{}}_{t+1:t+h} ,y_{t+1:t+h})\}_{t=w}^{n+w-1}$ is denoted by $L_{q}(\theta)$, and
\begin{equation}
L_{q}(\theta) =\sum _{t=w}^{n+w-1}\sum _{j=1}^{| \mathcal{Q}| }\sum _{i=1}^{h} \ell \left(\hat{y}_{t+i}^{( p_{j})} ,y_{t+i} ,p_{j}\right),
\end{equation}
where $\mathcal{Q}$ denotes a predetermined set of quantiles and $p_{j}$ is an element of $\mathcal{Q}$.

The performance of developed models was compared through two aspects: computational cost and prediction accuracy. The computational cost was expressed as the training time of each model in seconds. As suggested by the ASHRAE Guideline 14-2014~\cite{ashrae_guideline_14_measurement_2014}, prediction accuracy of point forecasting models was assessed by two scale-independent metrics, namely coefficient of variation of the root mean square error (CV-RMSE) and normalized mean bias error (NMBE), over the entire test set. They are calculated by Eqs.~\ref{eqn:cv-rmse} and~\ref{eqn:nmbe}.
\begin{equation}
RMSE=\sqrt{\frac{1}{n}\sum\limits _{t=1}^{n}(\hat{y}_{t} -y_{t})^{2}}, \label{eqn:rmse}
\end{equation}
\begin{equation}
CV\textrm{-}RMSE=\frac{RMSE}{\overline{y}} \times 100, \label{eqn:cv-rmse}
\end{equation}
\begin{equation}
MBE=\frac{1}{n}\sum\limits _{t=1}^{n}(\hat{y}_{t} -y_{t}), \label{eqn:mbe}
\end{equation}
\begin{equation}
NMBE=\frac{MBE}{\overline{y}} \times 100,  \label{eqn:nmbe}
\end{equation}
where $n$ is the length of forecast horizon, $y_{t}$ is the actual value of a target variable at time $t$, $\hat{y}_{t}$ is the predicted value of the target variable at time $t$, and $\overline{y}$ is the mean actual value of the target variable across the forecast horizon.

The CV-RMSE measures the variation between the actual values and the predictions of a model~\cite{ashrae_guideline_14_measurement_2014}. NMBE normalizes the mean bias error and gives the global difference between the actual and predicted values~\cite{ramos_validation_2017}. A positive NMBE value means that the model over-predicts actual energy consumption, and a negative one means under-prediction. For both CV-RMSE and NMBE, a value closer to zero indicates improved prediction accuracy. When making comparisons, we mainly focused on the CV-RMSE if the NMBE of a model is within the required range. As suggested by the ASHRAE Guideline 14-2014~\cite{ashrae_guideline_14_measurement_2014}, an applicable predictive model for energy use of whole building should have a CV-RMSE $\leq$~30\% and an NMBE within $\pm$10\% when using hourly data for training models. 

As in studies~\cite{salinas_deepar_2020,lim_temporal_2021}, the $\rho$-risk, which normalizes quantile losses across the entire forecast horizon, was utilized for assessing prediction accuracy of probabilistic forecasting models. $\rho$-risk at $p$th quantile is calculated by
\begin{equation}
\rho\textrm{-}risk(p) =\frac{2\times \sum\limits _{t=1}^{n} \ell \left(\hat{y}_{t}^{(p)} ,y_{t} ,p\right)}{\sum\limits _{t=1}^{n} y_{t}}, \label{eqn:rho_risk}
\end{equation}
where $n$ is the length of forecast horizon, $y_{t}$ is the actual value of a target variable at time $t$, $\hat{y}_{t}^{(p)}$ denotes the predicted $p$th quantile value at time $t$, and $\ell \left(\hat{y}_{t}^{(p)} ,y_{t} ,p\right)$ is the $p$th quantile loss calculated by Eq.~\ref{eqn:quantile_loss}. 

\chapter{The Digitalization Framework} 
\label{cha:digitalization_framework}

The digitization framework proposed in this study is a holistic solution. It can be used to collect and process necessary data to provide applications such as indoor environment monitoring and energy consumption prediction. The framework facilitates seamless information exchange among stakeholders and objects of interest by integrating IoT devices, cloud computing, information representation, and data analytics. This integration establishes a unified operating picture that enables the development of creative applications.

\section{Framework Design}

\begin{figure*}[!htb]
\includegraphics[width=\textwidth]{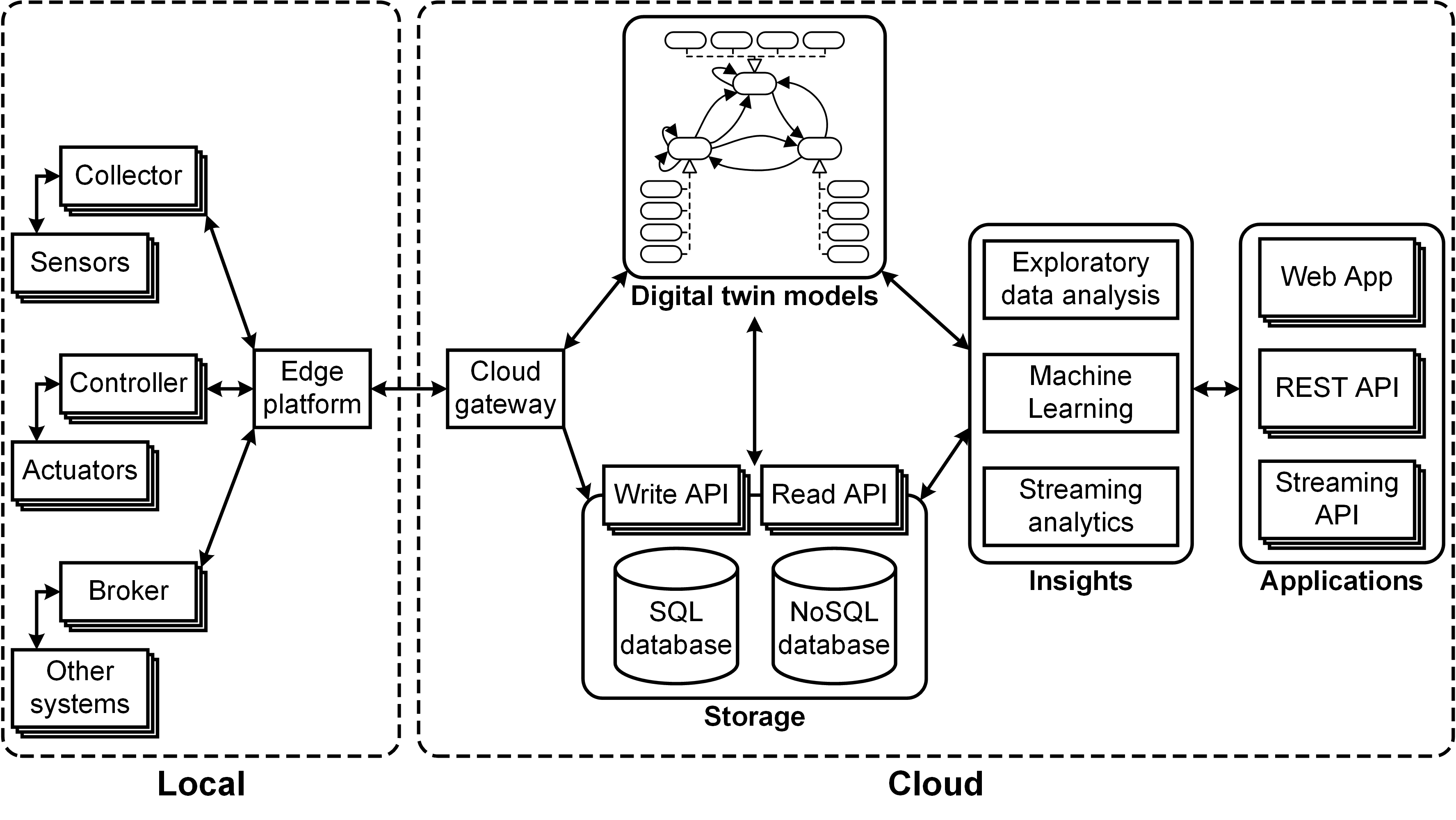}
\centering
\caption{The architecture of the proposed framework~\cite{ni_enabling_2022}.}\label{fig:framework_architecture}
\end{figure*}

As illustrated in Fig.~\ref{fig:framework_architecture}, the framework consists of the local part as well as the cloud part.

\subsection{The Local Part}

The local part comprises the edge platform and devices responsible for retrieving information from buildings. The devices are categorized into three groups based on their functions:
\begin{itemize}
    \item Sensors and collectors: Sensors measure parameters of interest, such as temperature, RH, and carbon dioxide (CO\textsubscript{2}). Collectors read the measurements from sensors.
    \item Actuators and controllers: Actuators are all the equipment that can be controlled, such as fans and heaters. An actuator receives commands from a controller and performs actions according to commands.
    \item Other systems and brokers: Other systems are those systems that exist in buildings before adopting our solution, such as BMS. Brokers facilitate retrieving data from these systems and sending instructions to them.
\end{itemize}

The edge platform has storage, computing, and networking resources. It has two main functions. First, it acts as a gateway, reporting data from local devices to the cloud and receives control instructions from the cloud. Second, it stores and processes data locally, taking on some AI and analytics workloads traditionally handled by the cloud platform. This local processing capability is beneficial for handling sensitive data or critical tasks that demand real-time performance because the data can be processed locally instead of being uploaded to the cloud. The edge platform communicates with local devices (collectors, controllers, and brokers) via wired or wireless networks, facilitating seamless data exchange and interaction.

\subsection{The Cloud Part}

\begin{itemize}
    \item Cloud gateway: The cloud gateway serves as the access point into the cloud. It manages edge platforms and allows for bidirectional communication between edge platforms and cloud. For instance, when a message arrives from an edge platform, the cloud gateway tells subsequent components to handle the message, e.g., storing relevant data and updating the status of digital twin models.
    \item Digital twin models: There are two stages to creating a parametric digital twin: (a) creating models of necessary physical entities and their relationships and (b) providing interfaces to access or update model status. Parametric digital twin models are constructed based on ontology, which is a consistent schema for data representation. Ontology facilitates the storage of metadata related to physical entities and captures the underlying relationships between them. By leveraging the ontology-based digital twin model, machine-readable data formats and query tools can be provided. These tools allow other modules to reason about the semantics of data.
    \item Storage: The storage includes databases for storing both structured and unstructured data. Other modules can interact with these databases using application programming interfaces (APIs). Structured data consists of telemetries gathered from sensors, actuators, and other well-organized data. Unstructured data include textual information and images. Examples of textual information include maintenance manuals for appliances or collections. Images could be periodic photographs that capture the condition of collections, such as paintings and sculptures, to monitor surface deterioration over time.
    \item Insights: Various data analytics methods can be used to uncover essential insights from historical and real-time data. Data analytics usually begins with EDA, which uses multiple techniques, notably graphical approaches, to understand a dataset thoroughly. Based on the insights supplied by EDA, machine learning methods can be utilized to develop predictive models that fulfill diverse purposes, such as anomaly detection and building energy forecasting. Once trained and validated, these predictive models are deployed to undertake streaming data analysis, allowing for real-time analysis.
    \item Applications: Applications can be in various forms, such as web applications as well as REST and streaming APIs. Examples of web applications include visualization tools for data and models or interactive simulation tools with features such as energy forecasting and occupancy prediction. APIs, on the other hand, function as contractual agreements between our systems and users of the information. For instance, an API developed for indoor environmental services may require users to provide identities (IDs) of a building and a room. In response, our system will provide users with two indoor environmental conditions: temperature and relative humidity.
\end{itemize}

A reference implementation of the digitalization framework can be found in \textbf{Papers III} and \textbf{IV}. The implementation is a comprehensive IoT system developed by ourselves to facilitate data collection with multiple environmental sensors. It also involves communicating data through edge platforms, storing data on the Microsoft Azure Cloud, as well as creating digital twins for historic buildings using ontology. The implementation focuses on modeling the indoor environment, thus enabling a better knowledge of conditions within buildings. Interfaces have been purposely reserved for future expansion, such as connecting equipment and other systems within the historic building to facilitate the exchange of operational status and control commands.

\section{Case Study}

\begin{figure}[!ht]
\includegraphics{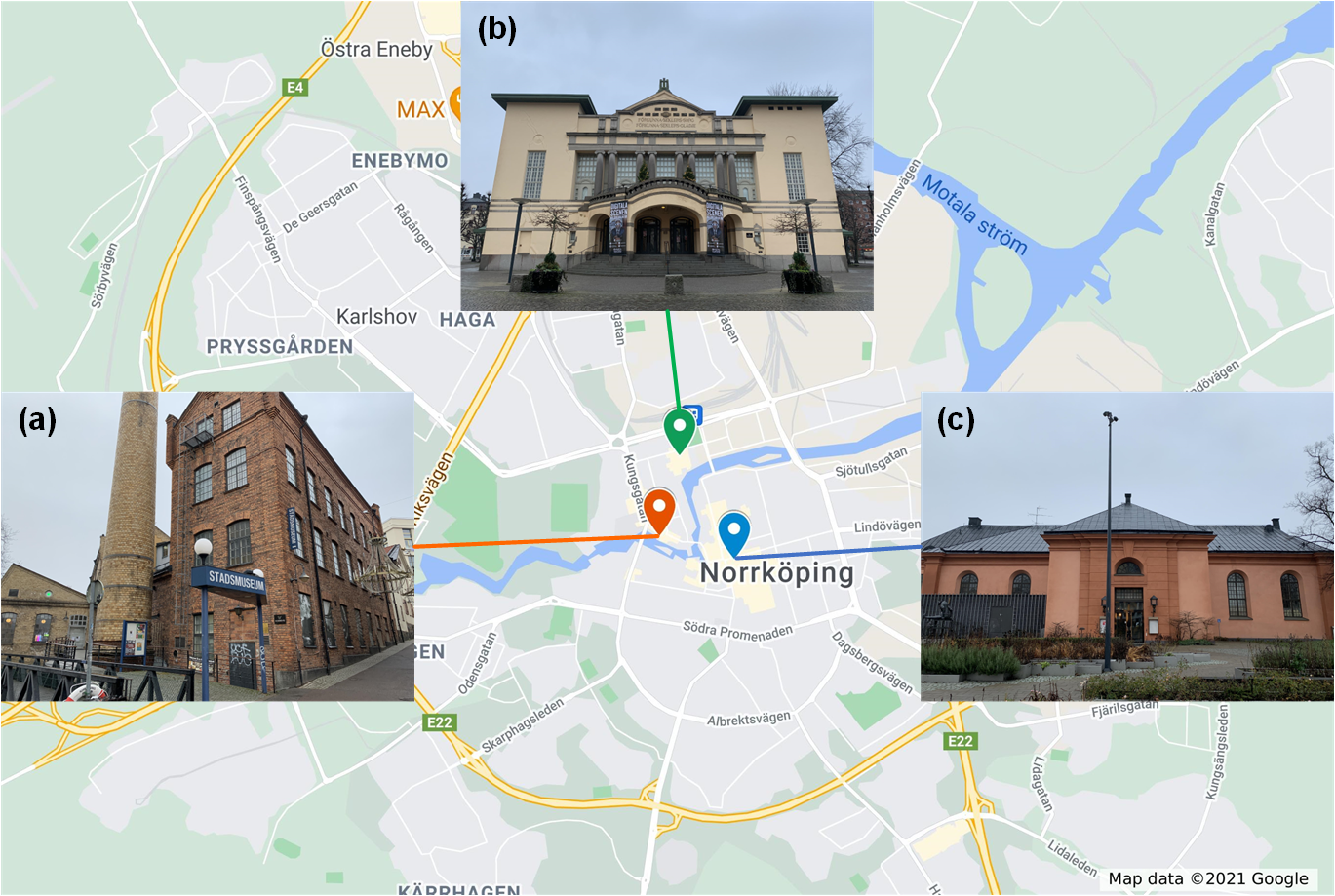}
\centering
\caption{Three case study buildings in Norrk\"oping, Sweden: (\textbf{a}) the City Museum, (\textbf{b}) the City Theatre, and (\textbf{c}) the Auditorium~\cite{ni_sensing_2021}.}\label{fig:case_study_buildings}
\end{figure}

The digitalization framework has been field tested in three historic buildings (as depicted in Fig.~\ref{fig:case_study_buildings}) situated in Norrk\"oping, Sweden. These buildings have a rich history spanning over a century and continue to serve as venues for diverse social and cultural activities. All of them have been listed as protected buildings. These buildings differ in construction age, internal structures, and current utilization. This deliberate selection enables us to evaluate the adaptability and portability of our solution in achieving digitalization objectives such as advanced monitoring and energy optimization with AI considering both heritage preservation and building functions.

\subsection{Description of Case Study Buildings}

The City Museum is housed in two properties that consist of five historic factory and craft buildings adjacent to the Motala River, within the old industrial landscape of Norrk\"oping~\cite{nkp_city_musuem_buildings_2023}. These buildings were built in the 19{th} and 20{th} centuries. The primary task of the City Museum is to convey the local cultural heritage, manage and care for the collections as well as gather and document new knowledge about the cultural heritage. The museum's extensive collection encompasses nearly 40,000 individual objects, including various items such as billboards, handicraft tools, printed fabrics, and sheets~\cite{nkp_city_musuem_collection_2023}. These objects serve as valuable historical artifacts and provide insight into the rich history of Norrk\"oping, particularly the development of crafts and industry in the 19{th} and 20{th} centuries.

The City Theatre was built in the Art Nouveau architectural style and was inaugurated in 1908~\cite{nkp_city_theatre_2023}. It is a venue for performing arts that reflect society and for meetings through cultural events. The City Theatre has a salon room that can seat 600 people. Prior to the COVID-19 pandemic, the theatre hosted over 50 shows per year in 2018 and 2019, attracting more than 13,000 visitors.

The Auditorium was built as a cruciform church in 1827~\cite{swedish_national_heritage_board_horsalen_2023}. The building was rebuilt in 1913 and served as the venue for the Norrk\"oping Symphony Orchestra until 1994~\cite{wiki_horsalen_2023}. Currently, the Auditorium is rented out for cultural activities such as concerts and lectures. The venue has 302 seats and four wheelchair spaces for accommodating up to 350 people simultaneously~\cite{norrkoping_horsalen_2023}. 

\subsection{Indoor Environment Monitoring}

Indoor environment monitoring of the three buildings began on March 16, 2021, and will continue until at least December 2023. Each case study building had one sensor box installed to collect six indoor environmental parameters: temperature, RH, CO\textsubscript{2}, particulate matter, harmful gas, and vibration. Each parameter is collected every 15~s. Such a high sampling rate ensures a low latency of the current status to capture impacts on the indoor environment, such as occupants' presence and unanticipated factors. Moreover, the high sampling rate provides high-resolution data that can be downsampled when indoor environment changes slowly. Undoubtedly, deploying more sensor boxes in each building can help us better study the spatial distribution of environmental parameters. Nevertheless, in this research, the installation position of each sensor box is typical. When more sensor boxes are placed, the methodology for creating digital twin models and the data analysis method are also applicable. Detailed analysis of indoor environment can be found in \textbf{Papers III} and \textbf{IV}.
 
\subsection{Building Energy Forecasting}

Both one-step-ahead (\textbf{Paper VI}) and multi-horizon (\textbf{Paper V}) point and probabilistic forecasting are studied.

\subsubsection{Data Collection}

The collected data have two parts. The first part is historical electricity use and heating load from the three case study buildings. They were provided by an energy engineer of the buildings. Both types of energy use data are of the entire building. The second part is historical weather data of Norrk\"oping, including dry-bulb temperature, relative humidity, dew point temperature, precipitation, wind speed, air pressure, and global irradiance. The weather data was obtained through open APIs~\cite{smhi_open_data_api_2023} provided by the Swedish Meteorological and Hydrological Institute (SMHI). All energy use and weather data are hourly data, spanning from 01:00 on January 1, 2016, to 00:00 on January 1, 2020. The time span of the collected data is before the pandemic of COVID-19, thereby excluding any potential impact of the pandemic on public activities held in these buildings.

\subsubsection{Data Pre-processing}

First, missing values in weather data were interpolated linearly. Then, the dataset was divided into three subsets: a training set (38 months, from January 1, 2016 to February 28, 2019) for learning the parameters of models, a validation set (five months, from March 1, 2019 to July 31, 2019) for tuning hyperparameters and preventing overfitting, and a test set (five months, from August 1, 2019 to December 31, 2019) for evaluating the performance of models. The dataset splitting roughly follows the empirical ratio of 80:10:10~\cite{gholamy_relation_2018} to achieve a good balance between overfitting and underfitting. The three subsets do not overlap in time, avoiding information leakage from the future. In addition, the training data covers multiple calendar years, which helps to provide energy use patterns of these buildings.

\subsubsection{Feature Engineering}

Feature engineering includes extracting temporal features from timestamps, generating features from operating modes of buildings, and reducing redundant features. Four temporal features are extracted: two binary and two cyclical variables. The binary variables are (1) \textit{is holiday} to indicate if a day is a Swedish public holiday and (2) \textit{is weekend} to indicate if a day is a weekend. The cyclical variables are \textit{hour} and \textit{weekday}. In addition to temporal features, one feature called \textit{is open} (binary variable) is added to reflect the occupancy of a building for a given hour.

A filter method based on finding the correlation between variables was employed to select critical features and reduce redundant features. The Pearson correlation coefficient ($r$) was used for measuring the linear relationship between two variables. As general rules of thumb, a threshold of $|r|\geq 0.3$ was adopted to filter out critical features with at least moderate correlation with a target variable. To reduce redundant features, when two features are highly correlated ($|r|\geq 0.7$), the one holding larger $|r|$ with the target variable was kept to avoid duplicate information.

Data transformation aims to change raw features into a more suitable representation for model learning. For electricity use and heating load, as well as weather data such as dry-bulb temperature and relative humidity, a min-max normalization was performed to scale each of them to a range of $[0, 1]$. All min-max scalers were fitted on the training set, then used for transforming validation and test sets. Cyclical features \textit{hour} and \textit{weekday} were transformed into two dimensions using a sine-cosine transformation. Binary features like \textit{is open} were not transformed.

\subsubsection{Studied Machine Learning Methods}

In addition to linear regression (LR), seven deep learning methods, namely hierarchical interpolation for time series forecasting (N-HiTS)~\cite{challu_n-hits_2022}, temporal convolutional network (TCN)~\cite{bai_empirical_2018}, Transformer (TF)~\cite{vaswani_attention_2017}, NLinear~\cite{zeng_transformers_2022}, long short-term memory (LSTM)~\cite{hochreiter_long_1997}, gated recurrent unit (GRU)~\cite{cho_properties_2014}, and temporal fusion transformer (TFT)~\cite{lim_temporal_2021}, were investigated to develop predictive models and compare their performance. N-HiTS and NLinear were improved to support producing probabilistic forecasts based on quantile regression. 

\chapter{Summary of Included Papers} 
\label{cha:summary_of_included_papers}

This chapter summarizes the included papers and highlights the author's contributions.

\section{Paper I}

\textbf{Z. Ni}, P. Eriksson, Y. Liu, M. Karlsson, and S. Gong, “Improving Energy Efficiency while Preserving Historic Buildings with Digital Twins and Artificial Intelligence,” \textit{SBE21 Sustainable Built Heritage}, Bolzano-Bozen, Italy, 14--16 Apr. 2021.

\textbf{Summary}: This paper proposes a methodology to develop a cloud-based solution for energy optimization and conservation of historic buildings. The solution integrates advanced digital technologies, such as IoT, cloud computing, and AI, to create digital twins of historic buildings. With collected data, data analytics is performed to improve maintenance of historic buildings to optimize energy use while still achieving conservation purposes. The paper also identifies research questions that need to be studied in energy optimization and building conservation.

\textbf{The author's contributions}: 1) Packaged sensors, a microcontroller, and a microprocessor in a plastic box for easy deployment. 2) Wrote software codes for the local part of the system to periodically read data from each sensor and encapsulate readings into messages to upload to Microsoft Azure IoT Hub. 3) Designed a cloud database and wrote a Microsoft Azure Function App to route messages to the database as well as tested the whole system. 4) Wrote most part of the initial manuscript draft and revised it according to feedback from co-authors.

\section{Paper II}

\textbf{Z. Ni}, Y. Liu, M. Karlsson, and S. Gong, “Link Historic Buildings to Cloud with Internet of Things and Digital Twins,” \textit{The 4th International Conference on Energy Efficiency in Historic Buildings}, Benediktbeuern, Germany, 4--5 May 2022.

\textbf{Summary}: This paper presents a reference implementation of a digitalization framework for energy optimization and building conservation. The implementation consists of two parts: local and cloud. The local part comprises multiple environmental sensors, edge devices, and communication hardware. The cloud part is built with several Microsoft Azure Cloud services. Functional validation in the laboratory has demonstrated the usefulness of the implementation.

\textbf{The author's contributions}: 1) Packaged sensors, a microcontroller, and a microprocessor in a plastic box for easy deployment. 2) Wrote software codes for the local part of the system to periodically read data from each sensor and encapsulate readings into messages to upload to IoT Hub. 3) Designed a cloud database and wrote a Function App to route messages to the database as well as tested the whole system. 4) Wrote most part of the initial manuscript draft and revised it according to feedback from co-authors.

\section{Paper III}

\textbf{Z. Ni}, Y. Liu, M. Karlsson, and S. Gong, “A Sensing System Based on Public Cloud to Monitor Indoor Environment of Historic Buildings,” \textit{Sensors}, vol. 21, no. 16, p. 5266, 2021.

\textbf{Summary}: This paper presents a cloud-based IoT system for real-time indoor environment monitoring. The system was tested in three historic buildings in Norrk\"oping, Sweden. The results show that data can be reliably collected, transmitted, and stored in the cloud database. Preliminary analysis of relative humidity fluctuations identifies potential risks to the buildings and housed artifacts.

\textbf{The author's contributions}: 1) Improved the sensor box package according to feedback from co-authors. 2) Deployed three sensor boxes to case study buildings with Shaofang Gong. 3) Implemented a Web app to visualize collected data. 4) Wrote the initial manuscript draft, including data analysis and visualization, and revised it according to feedback from co-authors.

\section{Paper IV}

\textbf{Z. Ni}, Y. Liu, M. Karlsson, and S. Gong, “Enabling Preventive Conservation of Historic Buildings Through Cloud-based Digital Twins: A Case Study in the City Theatre, Norrk\"oping,” \textit{IEEE Access}, vol. 10, pp. 90924–90939, 2022.

\textbf{Summary}: This paper discusses the importance of preventive conservation of historic buildings and the challenges in creating digital twins of such buildings. To address these challenges, a solution is proposed to combine IoT and ontology to create parametric digital twins that consistently represent the data and reflect the most recent conditions. The paper also gives a reference implementation that employs hardware, open-source software libraries, and the public cloud to reduce reinvention and make it easier to replicate and reuse in other historic buildings. A practical case study conducted in the City Theatre, Norrk\"oping, Sweden, illustrates the functionality and benefits of the created digital twin.

\textbf{The author's contributions}: 1) Involved ontology to create parametric digital twins in the Microsoft Azure Cloud. 2) Designed the method for qualitatively and quantitatively studying the impact of occupants’ presence on the indoor climate of the case study building. 3) Wrote the initial manuscript draft, including data analysis and visualization, and revised it according to feedback from co-authors.

\section{Paper V}

\textbf{Z. Ni}, C. Zhang, M. Karlsson, and S. Gong, “A Study of Deep Learning-based Multi-horizon Building Energy Forecasting,” Manuscript, 2023.

\textbf{Summary}: This paper has adapted and applied cutting-edge deep learning methods to the challenge of multi-horizon building energy forecasting. Eight different methods, including seven deep learning-based ones, namely N-HiTS, TCN, Transformer, NLinear, LSTM, GRU, and TFT, were studied to develop models for two public historic buildings in Norrk\"oping, Sweden, to make both point and probabilistic predictions. The two buildings have different use purposes. The performance of the developed models was examined, and the prediction under diverse energy consumption situations was investigated. The findings indicate that adding future knowledge on external factors influencing energy use is crucial for reliably multi-horizon forecasting. Furthermore, changes in the operating mode of a building and activities held in it increase uncertainty in energy use and reduce model prediction accuracy. The TFT model is the most competitive in point and probabilistic forecasting.

\textbf{The author's contributions}: 1) Conceived and designed the study. 2) Collected the weather data through APIs provided by the SMHI, performed data preprocessing, developed all predictive models, and conducted the model training and evaluation. 3) Wrote the initial manuscript draft, including data analysis and visualization, and revised it according to feedback from co-authors. 

\section{Paper VI}

\textbf{Z. Ni}, C. Zhang, M. Karlsson, and S. Gong, “Leveraging Deep Learning and Digital Twins to Improve Energy Performance of Buildings,” accepted in \textit{The 3rd IEEE International Conference on Industrial Electronics for Sustainable Energy Systems}, 2023.

\textbf{Summary}: This study introduces Deep Energy Twin, a solution that combines deep learning with digital twins to analyze building energy use and identify optimization opportunities. Ontology was employed to create parametric digital twins to offer a consistent data format for representing data across various systems in a building. Deep learning methods were utilized for data analytics. The conducted case study demonstrates the ability of five deep learning methods to predict building energy use and measure uncertainties. The findings suggest that deep learning methods are effective at capturing tendency and uncertainty in building energy use. This solution can give facility managers a greater understanding of how much energy is being used in their buildings, resulting in cost savings, improved human comfort, and a more sustainable built environment.

\textbf{The author's contributions}: 1) Conceived and designed the study. 2) Collected the weather data through APIs provided by the SMHI, performed data preprocessing, developed all predictive models, and conducted the model training and evaluation. 3) Wrote the initial manuscript draft, including data analysis and visualization, and revised it according to feedback from co-authors.

\chapter{Conclusion and Future Work} 
\label{cha:conclusion_and_future_work}

This thesis presents a comprehensive solution for smart maintenance of historic buildings. The solution entails the implementation of a digitalization framework that integrates various cutting-edge digital technologies, including IoT, cloud computing, ontology, and AI. By harnessing IoT, the solution enables the real-time collection of crucial operating data from historic buildings. Adopting ontology ensures consistent data representation across different systems within a historic building to create parametric digital twins. Furthermore, deep learning methods were utilized to analyze building energy use, identifying patterns and energy-saving opportunities. The main contributions of this thesis are summarized as follows.
\begin{itemize}
\item Proposed a cloud-based digitalization framework that integrates advanced digital technologies, such as IoT, cloud computing, ontology, and AI, for achieving smart maintenance of historic buildings. The cloud-based approach is estimated to bring $\sim$20\% cost savings~\cite{microsoft_trend_2017}. 
\item Presented a reference implementation using easy-to-use hardware, open-source software libraries, and the public cloud, i.e., Microsoft Azure Cloud, to minimize reinvention and make it easier to reproduce and reuse in other historic buildings. Furthermore, the reference implementation can be packaged as software as a service, which eases the delivery of applications and can be promoted to other historic buildings.
\item Validated and demonstrated the effectiveness of the solution by conducting a practical case study through two representative smart applications in three public historic buildings with different use purposes located in Norrk\"oping, Sweden. The findings obtained from indoor environment monitoring and building energy forecasting brought valuable insights not only for the maintainer of case study buildings in this study but also for maintainers of other historic buildings with similar use purposes.
\end{itemize}

In the future, it is worth further investigating the application of AI in smart maintenance of historic buildings. Digital twin models that are integrated with powerful AI could enable comprehensive simulations of building operating status under various conditions. By leveraging these simulations, areas requiring maintenance optimization can be precisely identified, thereby enhancing overall sustainability and efficiency in preserving historic buildings.


